\documentclass[runningheads]{llncs}

\usepackage[T1]{fontenc}

\usepackage{graphicx}
\usepackage{cite}
\usepackage{amsmath,amssymb,amsfonts}
\usepackage{algorithmic}
\usepackage{graphicx}
\usepackage{textcomp}
\usepackage{xcolor}
\usepackage{subcaption}
\usepackage{pgfplots}

\newcommand{\SYS}{LNSP}

\begin{document}
\begin{sloppypar}   

\title{Offline Map Matching Based on Localization Error Distribution Modeling}
\titlerunning{Offline Map Matching for Sparse Trajectories Based on LED Modeling}

\author{Ruilin Xu\inst{1}\orcidID{0009-0006-9320-8012} \and
Yuchen Song\inst{2}\and
Kaijie Li\inst{2}\orcidID{0009-0000-1697-3323}\and
Xitong Gao\inst{2}\orcidID{0000-0002-2063-2051}\and
Kejiang Ye\inst{2}\orcidID{0000-0001-6133-407X}\and
Fan Zhang  \inst{2}\and
Juanjuan Zhao\inst{2}\thanks{Corresponding author: Juanjuan Zhao, email: jj.zhao@siat.ac.cn}\orcidID{0000-0003-1002-9272}}

\authorrunning{R. Xu et al.}

\institute{{Southern University of Science and Technology, Shenzhen, China 
\email{xurl2024@mail.sustech.edu.cn}}\\\and
Shenzhen Institute of Advanced Technology, Chinese Academy of Sciences, Shenzhen, China
\email{\{yc.song, kj.li,  xt.gao, kj.ye, jj.zhao\}@siat.ac.cn}}

\maketitle              
\vspace{-0.5cm}
\begin{abstract}

Offline map matching involves aligning historical trajectories of mobile objects, which may have positional errors, with digital maps. This is essential for applications in intelligent transportation systems (ITS), such as route analysis and traffic pattern mining. Existing methods have two main limitations: (i) they assume a uniform Localization Error Distribution (LED) across urban areas, neglecting environmental factors that lead to suboptimal path search ranges, and (ii) they struggle to efficiently handle local non-shortest paths and detours. To address these issues, we propose a novel offline map matching method for sparse trajectories, called {\SYS}, which integrates LED modeling and non-shortest path detection. Key innovations include: (i) leveraging public transit trajectories with fixed routes to model LED in finer detail across different city regions, optimizing path search ranges, and (ii) scoring paths using sub-region dependency LED and a sliding window, which reduces global map matching errors. Experimental results using real-world bus and taxi trajectory datasets demonstrate that the {\SYS} algorithm significantly outperforms existing methods in both efficiency and matching accuracy.

\keywords{GPS error distribution  \and Map matching \and Hybrid trajectories}

\end{abstract}

\vspace{-1cm}
\section{Introduction}

Offline map matching (MM) aligns historical trajectory data with digital maps, correcting GPS errors from signal reflection and multipath interference. It is crucial for ITS, aiding in route planning and traffic pattern analysis.
Trajectory sampling frequency is often reduced to save storage and communication costs. However, lower sampling rates increase ambiguity in route identification between GPS points. Current methods use dynamic programming algorithms based on local shortest paths for MM~\cite{RAHMANI201341}, but they face limitations due to poor LED modeling and inaccurate path scoring.

1) LED modeling: LED affects how paths are searched, but current MM methods often assume LED is the same across all urban areas, ignoring differences caused by factors like mountains and buildings. To avoid missing the correct road, the search radius is often set wide. However, as the number of trajectory points increases (e.g., to hundreds), the search space becomes too large, significantly reducing matching efficiency.

2) Path scoring: Current methods primarily rely on the geographical distance between trajectory points and candidate path to get the matching score. However a closer between the location point and the candidate road does not necessarily indicate a higher match accuracy since the different LED. This leads to inaccuracies in path scoring, especially when trajectory points are sparse, or the vehicle does not follow the shortest path.

To address these limitations, this paper integrates bus trajectories, which have fixed routes, and proposes an offline MM method that accurately models LED and detects non-shortest paths. The main contributions include:

1) It uses bus trajectories with fixed routes to capture localization error patterns in fine-grained urban areas. To address the uneven coverage of bus trajectories, hierarchical spectral clustering is applied to group city grids, identifying sub-regions composed of geographically adjacent grids with similar LED.

2)  It improves the robustness and efficiency of map matching by adjusting the search radius of candidate routes, scoring each route based on region-dependent LEDs, and connecting local sub-trajectories using overlapping sliding windows.
 
3) Experiments conducted  on real trajectory dataset of buses and taxis in Shenzhen, China for one month show that our method significantly outperforms existing mainstream methods in both map matching efficiency and accuracy.

\vspace{-0.1cm}
\section{Related Work}

Offline Map Matching  is designed to precisely match historical and complete trajectory data with electronic maps. It is crucial for vehicle trajectory analysis, traffic pattern mining, and trajectory compression~\cite{article4}. 

Some offline matching algorithms assume global  point correlation of a trajectory based on the shortest path. Srivatsa et al.~\cite{Srivatsa2013MapMF} identify candidate paths using the shortest distances from all trajectory points to eahc candidate path as the weight. Rahmani et al.~\cite{RAHMANI201341} consider both path length and match with travel time in calculating path weight.~\cite{lou2009map, newson2009hidden, giovannini2011novel, zheng2012reducing} also simultaneously consider positions and time factors between adjacent GPS trajectory points, transforming the matching process into an optimization problem where the optimal solution corresponds to the matched roads. For example, in ~\cite{lou2009map}, candidate road segments are organized into a matching graph based on adjacent-time trajectory points, and the best matching path is derived from this graph, representing the matching result. Similarly, in ~\cite{giovannini2011novel}, a voting mechanism, Hidden Markov Model, and A* algorithm are employed to solve the optimization model. Quddus et al.~\cite{QUDDUS2015328} include distance weights and heading differences.
Since the shortest path is not always followed in real scenarios~\cite{10.1371/journal.pone.0134322}, Zhu et al.~\cite{6b4e4824cdd5492bb9cfec5042c7f614} proposed a trajectory segmentation optimization model using dynamic programming to align trajectory and path curves. Chambers et al.~\cite{10.1145/3191801.3191812} break down the global path into k local shortest paths, using the Fréchet distance between trajectory and path curves as the weight.

Existing methods use a larger search radius to tolerate positioning errors, which can include irrelevant roads and increase the matching burden~\cite{8344804}.

\vspace{-0.3cm}
\section{Overview}
\vspace{-0.1cm}
\label{sec:overview}
 \subsection{Terms and Problem Definition}
\vspace{-0.1cm}
\vspace{-1cm}
\begin{table}[ht]
\centering
\caption{Definitions of Terms}
\begin{tabular}{c|p{9cm}}
\hline
\textbf{Term} & \textbf{Definition} \\
\hline
GPS Point & A GPS point \( z = (lon, lat, t) \) is the position of a vehicle, consisting of longitude, latitude, and timestamp. \\
\hline
Trajectory & A trajectory is an ordered sequence of GPS points \( Z = (z_1, z_2, \dots, z_{|Z|}) \) collected by a moving vehicle, where \( z_1 \) is the start, \( z_{|Z|} \) is the end, and \( z_i.t \leq z_{i+1}.t \) for all \( i \). \\
\hline
Road Network & A directed graph \( \mathcal{G} = (\mathcal{V}, \mathcal{E}) \), where \( \mathcal{V} \) and \( \mathcal{E} \) is the set of intersections and directed road segments. A segment \( e \in \mathcal{E} \) is associated with an identifier \( id \), speed limit \( v \), length \( l \), start point \( start \), end point \( end \), and intermediate points. \\
\hline
Candidate Road & A candidate road is a road segment within a specified radius of a GPS point, where the point may potentially match. \\
\hline
Candidate Matched Point & It is the closest position on a candidate road to a target GPS point, denoted as \( s \). The road and the error is is \( s.c \)  and \( s.err \). \\
\hline
Path & A path between two GPS points \( p_{\text{start}} \) and \( p_{\text{end}} \) is a sequence of road segments \( P_{p_{\text{start}} \to p_{\text{end}}} = e_1 \to e_2 \to \dots \to e_n \), where \( e_1 \) and \( e_n \) correspond to the segments containing \( p_{\text{start}} \) and \( p_{\text{end}} \), and \( e_k.end = e_{k+1}.start \). \\
\hline
\end{tabular}

\end{table}

\begin{figure*}[htbp]
\vspace{-1.3cm}
\centering
\includegraphics[width=0.9\textwidth]{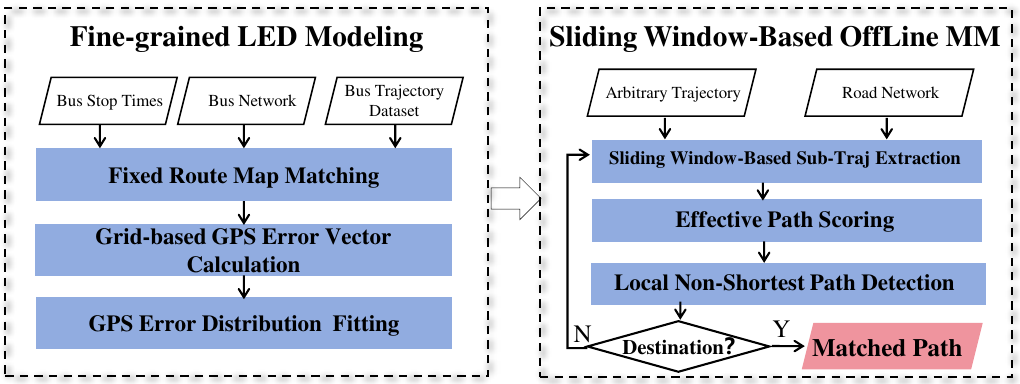}
\caption{{\SYS} Framework}
\label{fig:framework}
\vspace{-0.5cm}
\end{figure*}
\vspace{-0.4cm}
\textbf{Problem Definition}:  Given a trajectory \( Z \) and a road network \( G \), the goal is to estimate the actual travel path \( P \) that matches \( Z \).

\vspace{-0.3cm}
\subsection{Framework}
\vspace{-0.1cm}

The {\SYS} framework, shown in Fig.~\ref{fig:framework}, consists of two main components: Fine-grained LED Modeling and Sliding Window-Based Offline MM. Fine-grained LED Modeling integrates bus travel paths, maps historical trajectories to roads, and constructs a detailed LED function for urban areas. The Sliding Window-Based Offline MM enhances offline map matching efficiency and accuracy by adjusting the search radius for candidate roads, scoring paths using LED, and linking sub-trajectories with overlapping windows.

\vspace{-0.2cm}
\section{Methodology}

\subsection{Fine-grained LED Modeling}
\vspace{-0.2cm}
This section leverages  bus trajectories with  fixed routes to capture the fine-grained LED. 
To improve efficiency and accuracy, we segment bus trajectories using the actual arrival and departure times of the bus at each station, narrowing the search scope.

The city area is divided into equal-sized grids, denoted as \(\{g_1, g_2, \dots, g_M\}\), \(M\) is the total number of grids. Bus trajectory points are mapped into these grids, and a GPS error histogram vector \(\vec{V}_j\) is created for each grid \(g_j\), where each element represents a probability distribution for a specific error interval. To address uneven GPS point distribution and sparse data in some grids, we use the spatial similarity of nearby grids. Graph message passing is applied to fill in missing data by averaging error histograms from adjacent grids. Then, spectral clustering is used to group neighboring grids with similar GPS error patterns into same sub-region. 

The similarity of localization error distributions between two grids \(g_i\) and \(g_j\) is calculated as followings: For adjacent grids, we calculate similarity using a Gaussian kernel:
$W_{ij} = \exp \left( -\frac{\| \vec{V}_i - \vec{V}_j \|^2}{2\sigma^2} \right)$, where \(\vec{V}_i\) and \(\vec{V}_j\) are the error histogram vectors for grids \(g_i\) and \(g_j\), and \(\sigma\) is the Gaussian kernel's bandwidth parameter.  For non-adjacent grids, the similarity is set to 0 to keep the cells of same sub-region connected. Therefore, We  create \(K_g\) subregions \(\{\Gamma_1, \Gamma_2, \ldots, \Gamma_{K_g}\}\). Each subregion consists of connected grid cells, and each grid is assigned a subregion identifier, stored in an array that maps grid indices to their respective subregions.

Finally, a LED function is fitted for each subregion ${{g}_{j}}$. Based on the vector ${{\vec{V}}_{j}}$, we select an appropriate probability distribution ${{U}_{i}}$ (e.g., Gaussian, mixture of Gaussians, exponential, or log-normal) and fit it iteratively. The parameters are determined using Maximum Likelihood Estimation (MLE), and the fit's effectiveness is evaluated using the Akaike Information Criterion (AIC~\cite{1100705}). The cumulative distribution function as ${{A}_{i}}$. For these subregions that don't fit any common distributions, we use the error histogram vector to represent them.

\vspace{-0.5cm}
\subsection{Sliding Window-Based OffLine MM}

\subsubsection{The basic processing flow} 
\begin{figure}[htbp]
\vspace{-0.4cm}
\centering
\includegraphics[width=0.7\textwidth]{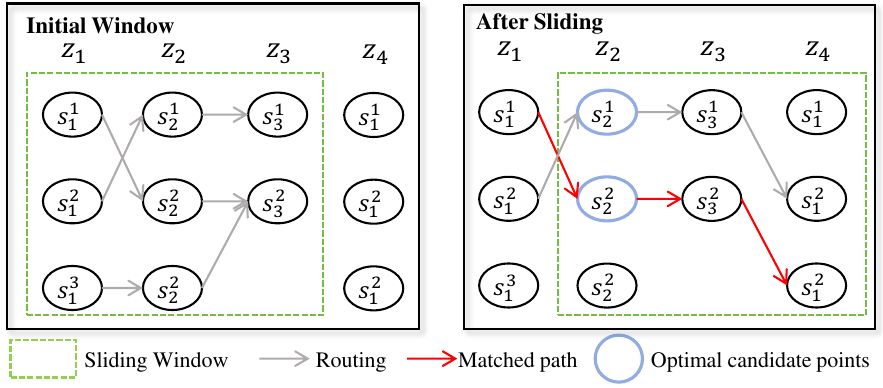}
\caption{Sliding Window Matching Process}
\vspace{-0.8cm}
\label{fig:sliding_window}
\end{figure}

Given a trajectory \( Z = (z_1, z_2, \dots, z_n) \) and a fixed-length window of size \( W_{\text{len}} \), the window starts at the first point (\( w_{\text{left}} \)) and ends at the \( W_{\text{len}} \)-th point (\( w_{\text{right}} \)), as shown in Fig.~\ref{fig:sliding_window}. For the local trajectory within this window, we first identify the grids where the starting point \( z_{w_{\text{left}}} \) and ending point \( z_{w_{\text{right}}} \) are located, along with their corresponding LED functions. GPS errors that exceed a certain threshold (e.g., 99\% cumulative probability) define the search radius for candidate points. This gives us two sets of candidate points: 
- \( S_{w_{\text{left}}} = (s_{w_{\text{left}}}^1, s_{w_{\text{left}}}^2, \dots, s_{w_{\text{left}}}^m) \) for the start points, and
- \( S_{w_{\text{right}}} = (s_{w_{\text{right}}}^1, s_{w_{\text{right}}}^2, \dots, s_{w_{\text{right}}}^n) \) for the end points.  Next, Dijkstra's algorithm is used to find the shortest path \( P_{s_{w_{\text{left}}}^i \to s_{w_{\text{right}}}^j} \) between each pair of start and end candidates, resulting in up to \( m \times n \) possible path segments. For each shortest path, a fixed path matching method is applied, and the path score \( P_{s_{w_{\text{left}}}^i \to s_{w_{\text{right}}}^j}.f \) is computed, as described in the next section. Finally, we select the top \( k \) paths with the highest scores.

The window is then slid backward by a fixed length ${{W}_{len}}$, ensuring the overlap between windows is at least ${{W}_{olen}}$ (as shown by the green window in Fig.~\ref{fig:sliding_window}). The values of ${{w}_{left}}$ and ${{w}_{right}}$ are updated accordingly. The starting point in the new window is now ${{z}_{{{w}_{left}}}}$, and the candidate points are defined as the points $P.{{s}_{{{w}_{left}}}}$ corresponding to ${{z}_{{{w}_{left}}}}$ in the retained paths from the previous window (illustrated by the blue-marked points in Fig.~\ref{fig:sliding_window}, with ${{W}_{olen}}=2$). These candidate points are considered more optimal for positioning ${{z}_{{{w}_{left}}}}$ based on the previous window. Each candidate point $s_{{{w}_{left}}}^{j}$ retains its associated path $s_{{{w}_{left}}}^{j}.P$ from the previous window, as well as its predecessor point $s_{{{w}_{left}}}^{j}.pres$ and predecessor route $s_{{{w}_{left}}}^{j}.prer$. 

Using the same method as the initial window, candidate points for the endpoint are identified, and the shortest paths are computed. From the new window onward, each path gets a score and an accumulated score $P.af$, which combines the current score with the previous window's score at the starting point (in the initial window, $P.af$= $P.f$ ). The accumulated score $af$ for each path ${{P}_{s_{{{w}_{left}}}^{i}\to s_{{{w}_{right}}}^{j}}}$ in the new window is calculated as:

\begin{tiny}
\begin{equation}
{{P}_{s_{{{w}_{left}}}^{i}\to s_{{{w}_{right}}}^{j}}}.af=    
\left\{ \begin{array}{*{35}{l}}
   {{P}_{s_{{{w}_{left}}}^{i}\to s_{{{w}_{right}}}^{j}}}.f &  \text{ }{{w}_{left}}=1  \\
   {{P}_{s_{{{w}_{left}}}^{i}\to s_{{{w}_{right}}}^{j}}}.f+s_{{{w}_{left}}}^{j}.P.af &  \text{ }{{w}_{left}}>1  \\
\end{array} \right.
\end{equation}
\end{tiny}

Finally, the top $k$ paths with the highest $wf$ values are retained in the window. The window continues to slide backward, performing the same calculations.

Once the window slides to the end of the trajectory and the calculations are finished (Fig.~\ref{fig:sliding_window}), each retained path has a cumulative score \( wf \), representing the total weight of the path and its preceding routes. By backtracking from the endpoint of the path with the largest \( wf \), using the preserved predecessor routes, the final matched path is identified.

\vspace{-0.3cm}
\subsubsection{Path Scoring Based on Sub-Region LED}
Existing methods for calculating path weights rely on the spatial similarity between the trajectory and the path, assuming a uniform error distribution across all regions. This paper refines the scoring method. After matching the shortest path $P$ to the trajectory $Z=({{z}_{{{w}_{left}}}},{{z}_{{{w}_{left}}+1}},\ldots ,{{z}_{{{w}_{right}}}})$ and obtaining the set of matching points $S=({{s}_{{{w}_{left}}}},{{s}_{{{w}_{left}}+1}},\ldots ,{{s}_{{{w}_{right}}}})$, each matching point ${{s}_{i}}$ corresponding to a positioning point ${{z}_{i}}$ has a matching error value ${{s}_{i}}.err$. The probability ${{F}_{s{{r}_{{{s}_{i}}}}}}({{s}_{i}}.err)$ of ${{s}_{i}}$ matching ${{z}_{i}}$ is calculated based on the sub-region $s{{r}_{{{s}_{i}}}}$ where ${{s}_{i}}$ is located and its corresponding LED function ${{U}_{s{{r}_{{{s}_{i}}}}}}$. The path weight score $P$ is the sum of the probabilities of all matching points in $S$ by $P.f=\sum\limits_{i={{w}_{left}}}^{{{w}_{right}}}{{{F}_{s{{r}_{{{s}_{i}}}}}}}({{s}_{i}}.err)$
\vspace{-0.3cm}
\subsubsection{Detection of Local Non-Shortest Paths}

\begin{figure}[htbp]
\vspace{-0.8cm}
    \centering
    \begin{minipage}[b]{0.45\textwidth}
        \centering
        \includegraphics[width=\textwidth]{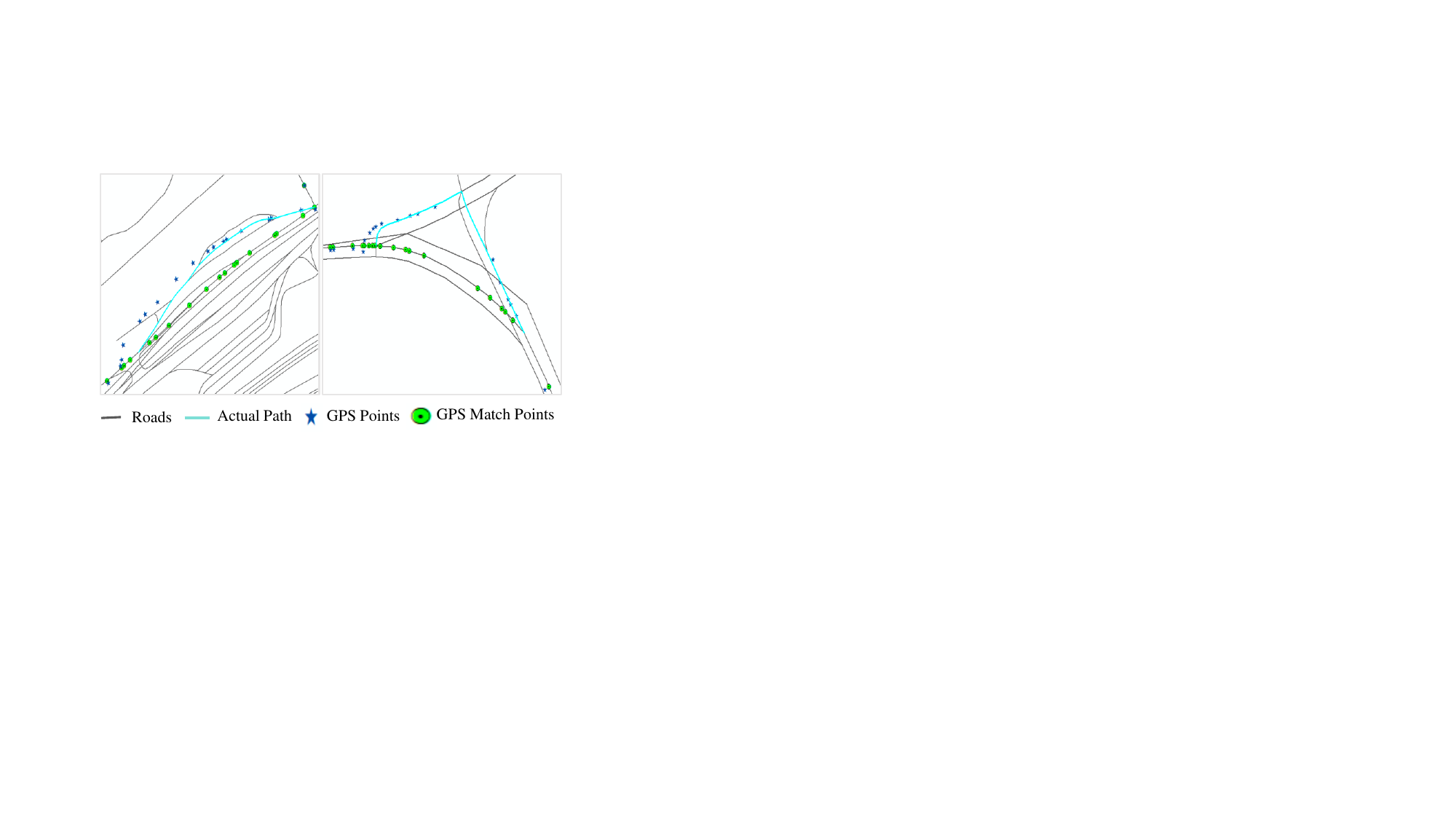}  
        \caption{Local Non-Shortest Path}
        \label{fig:non_shortest_path}
    \end{minipage}
    \hfill
    \begin{minipage}[b]{0.45\textwidth}
        \centering
        \includegraphics[width=\textwidth]{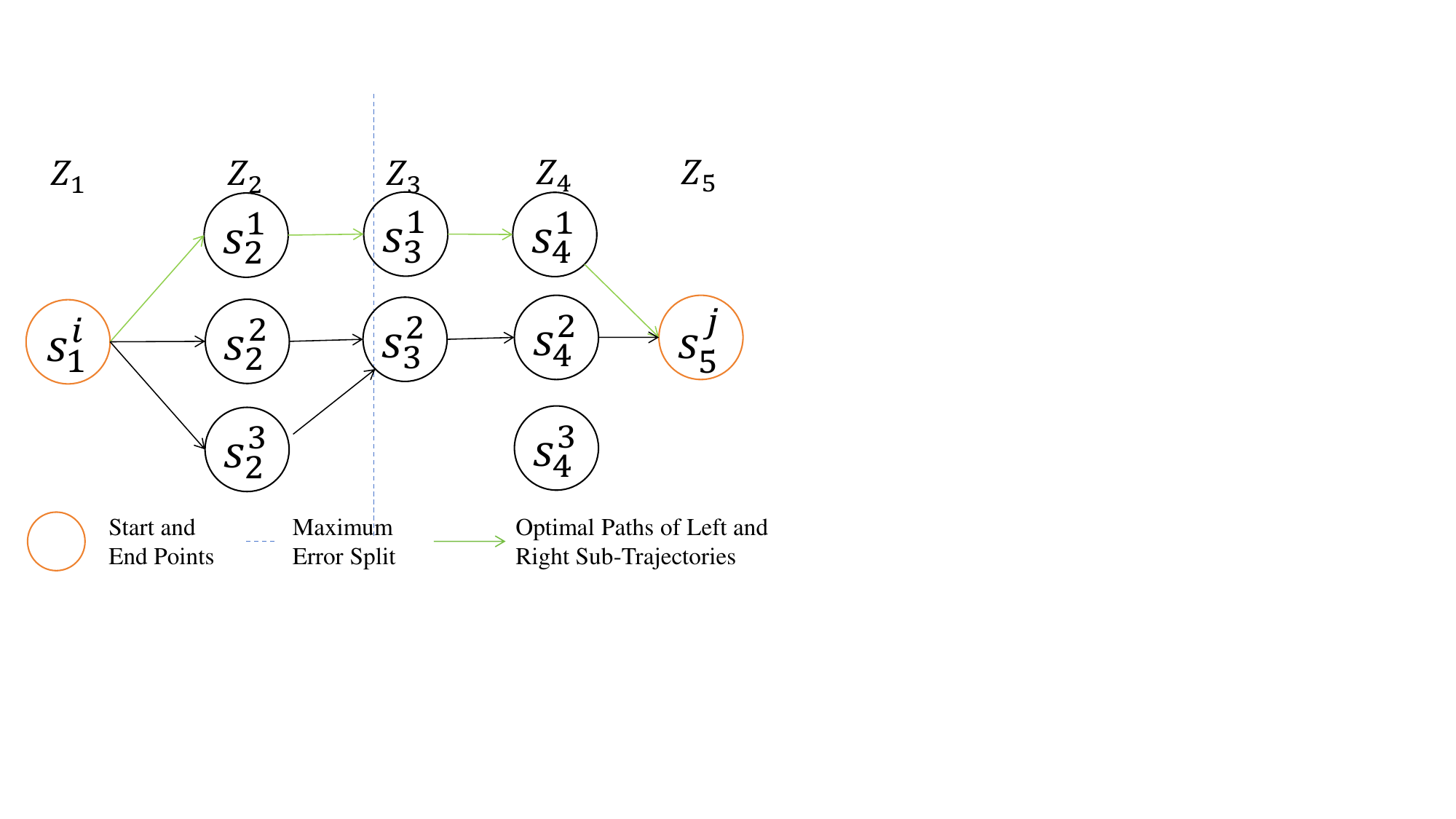} 
        \caption{Local NP Detection}
        \label{fig:segmentation_nonSP}
        \end{minipage}
    \vspace{-0.5cm}  
    \end{figure}

The sliding window method effectively handles global non-shortest path (NSP) driving, ensuring continuity in local shortest path matching and the correlation between consecutive trajectories. However, finding suitable window parameters for all trajectories is challenging, and shorter NSP segments can still occur in rare cases. For example, as shown in Fig.~\ref{fig:non_shortest_path}, a detour of less than 100 meters due to route regulations can cause the shortest path matching algorithm to fail. This paper analyzes the data and explores the geometric and topological characteristics of urban roads. It finds that when a vehicle deviates from the shortest path, it usually first moves away from the path and then returns. This incorrect matching is marked by a sequence of points with large errors, where the positioning points drift away and then gradually return, a common feature of detour trajectories.

Based on this characteristic, the paper proposes a method to detect local NSP segments. If a sequence of matching points contains a segment of \( NSP.n \) consecutive points with error distances greater than \( NSP.err \), where the maximum error exceeds \( NSP.\max\_err \), the segment length exceeds \( NSP.len \), and the error distribution significantly deviates, the segment is flagged as potentially containing a local NSP. The trajectory is then split at the point where the error exceeds \( NSP.\max\_err \) (see Fig.~\ref{fig:segmentation_nonSP}). The left half of the trajectory is recursively analyzed for shortest path matching, producing a list of possible path sequences, \( leftSequenceList \). The endpoints of each path in \( leftSequenceList \) are used as candidates for the right half of the trajectory, which is recursively matched with the fixed endpoint of the original sequence, resulting in \( rightSequenceList \). The path with the highest score is selected and combined with the best path from \( leftSequenceList \) to form a new matching path, \( nseq \). If \( nseq \) has a higher score than the original sequence, \( seq \) is replaced with \( nseq \).

This recursive method enhances performance by splitting the trajectory at the maximum error point into two shortest-path segments. It handles multiple local non-shortest path segments and terminates when the split segment length is below a threshold $NSP.len$ or if the optimal match matches the original sequence, avoiding excessive splits.
\vspace{-0.4cm}
\section{Experiment}
\vspace{-0.2cm}
\subsection{Experiment settings}
\vspace{-0.1cm}
\subsubsection{Dataset}
The used datasets include four types in Shenzhen, China: road networks, public transit routes, bus and taxi trajectories. In addition to analyzing GPS error distribution, the bus trajectory data is employed for testing, which assumes the paths are unknown.

\textbf{Road Network Data}: It is sourced from OpenStreetMap, focusing on the high-density road network in Futian District. The specific attributes of the road network are listed in Table~\ref{tab:futianroad}.

\begin{table}[htbp]
\vspace{-1cm}
\caption{Basic Attributes of Target Road Network}
\label{tab:futianroad}
\centering
\begin{tabular}{c|c|c|c|c|c}
\hline
\textbf{Attribute} & \textbf{Value} & \textbf{Attribute} & \textbf{Value} & \textbf{Attribute} & \textbf{Value} \\
\hline
Node Number & 2813 & Shortest Road & 4.5m & Road Number & 6892 \\
Road Degree & 2.45 & Road Length & 6.9m & Longest Road Length & 2910m \\
\hline
\end{tabular}
\vspace{-0.9cm}
\end{table}

\textbf{Bus Lines}: Each bus line is made up of a series of stops and the road segments it travels. These segments were collected by Shenzhen Transportation Department using high-precision GPS devices.

\textbf{Taxi Trajectories}: It spans from February 1 to February 28, 2023. The taxis' onboard devices record location every 5 to 30 seconds, including latitude, longitude, time, speed, direction, and passenger status. The location sequence is divided into individual trips (trajectories) by detecting transitions between passenger-carrying and non-carrying states. To obtain the actual path as the map matching ground truth for each trajectory, we selected only high-sampling-rate trajectories (with intervals of less than 3 seconds per point), ensuring good matching results even with conventional methods. The results from  dynamic programming algorithm-based map matching are used as the ground truth. For validation, we downsampled these trajectories to create low-sampling-rate trajectories (ranging from 5 seconds to 2 minutes) by randomly discarding some points. The closer the matching results under low sampling rates are to those under high sampling rates, the better the matching performance. 

\textbf{Bus Trajectories}: We use bus trajectories for the same period as the taxi data. Bus GPS devices record location every 5 seconds and provide additional details, such as bus line and direction (inbound or outbound). 

\vspace{-0.6cm}
\subsubsection{Metrics}
Matching accuracy and efficiency are critical performance metrics for map matching, and they are widely used in existing researches~\cite{Srivatsa2013MapMF,RAHMANI201341,QUDDUS2015328}. We evaluate the results using three metrics: Accuracy of Point Matching(Acc), Path Matching Precision (Prc), Average Matching Time (Mt).

\vspace{-8pt}
\[
\scriptsize
\text{Acc} = \frac{\text{Correct matches}}{\text{Total points}}, \quad 
\text{Prc} = \frac{\text{Length of correct paths}}{\text{Total path length}}, \quad 
\text{Mt} = \frac{\text{Matched trajectories}}{\text{Total matching time}}
\]

\subsubsection{Baselines}
We use three widely adopted Offline map matching algorithms as benchmarks: SPT~\cite{RAHMANI201341}, stMM~\cite{QUDDUS2015328}, and k-spMM~\cite{10.1145/3191801.3191812}.

\textbf{SPT}~\cite{RAHMANI201341}: is a global shortest-path based matching algorithm that considers path-trajectory length similarity, the Fréchet distance between trajectory and path curves, and the matching of global path length with travel time.  

\textbf{stMM}~\cite{QUDDUS2015328}: it adds two weights to the traditional curve distance and azimuth difference: the distance between the shortest path and vehicle trajectory, and the heading difference between the path and trajectory.  

\textbf{k-spMM}~\cite{10.1145/3191801.3191812}: is a local shortest-path matching algorithm that splits the global path into K local shortest paths using dynamic programming, with the Fréchet distance as the weight, aiming to minimize the overall matching cost.

\vspace{-0.6cm}
\subsubsection{Parameter Settings}
We set all parameters of sliding Window based on test result. The best setting is: Minimum trajectory length within the window ${W_{len}} = 600m$, minimum overlap length between adjacent sliding windows ${W_{olen}} = 300m$, and the number of paths retained within a single window $k = 4$.

\vspace{-0.4cm}
\subsection{Analysis of Experimental Results}
\vspace{-0.1cm}
\subsubsection{Region Division and Localization Error Distribution}

We divided the Futian District into 100-meter square grids. Spectral clustering is used to group the cells based on LED, resulting in 179 sub-regions. The average errors are shown in Table~\ref{tab:numbersubregions}. Most sub-regions have average errors between 0 and 10, with only a few having higher errors. This demonstrates that dynamically adjusting the candidate road search radius based on error distributions is effective, avoiding the need for a large fixed global search radius and validating our approach.

\begin{figure}[htbp]
\centering
\includegraphics[width=0.8\textwidth]{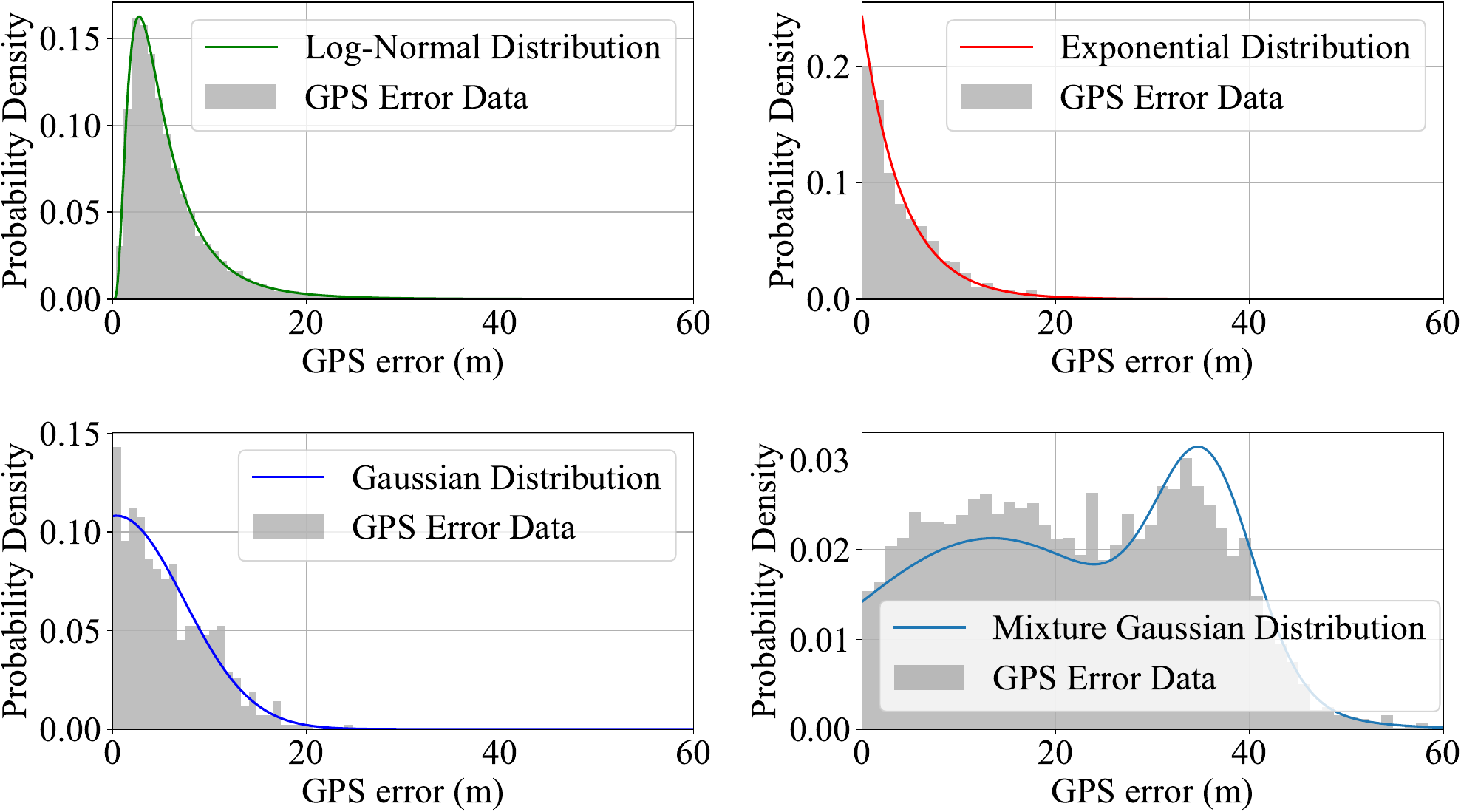}
\vspace{-0.3cm}
\caption{Examples of Sub-region GPS Error Distribution Fitting}
\label{fig:sub_region_distribution_fitting}
\vspace{-0.8cm}
\end{figure}

For the 179 divided regions, Table~\ref{tab:numbersub-regions} shows the LED functions and Figure.~\ref{fig:sub_region_distribution_fitting} provides some  examples. Among these, 54.1\% of the sub-regions' LED fit Gaussian or mixture of Gaussians distributions, while a small number fit log-normal or exponential distributions. For 24.5\% of the sub-regions that cannot be effectively fitted, GPS histogram vectors are used instead. 

\vspace{-1cm}
\begin{table}[htbp]

\setlength\tabcolsep{1pt}
\caption{Sub-Regions with Different Error Ranges}
\label{tab:numbersubregions}
\centering
\begin{tabular}{c c c c c c c}
\hline
\textbf{Average Error(m)} & [0, 5) & [5, 10) & [10, 15) & [15, 20) & [20, 25) & [25, 100) \\
\hline
\textbf{Number of Sub-Regions} & 42 & 93 & 24 & 12 & 10 & 8\\
\hline
\end{tabular}
\end{table}

\vspace{-1cm}
\begin{table}[htbp]
\vspace{-0.8cm}
\caption{Sub-Regions with Different LED Functions}
\label{tab:numbersub-regions}
\centering
\begin{tabular}{cccc}
\hline
\textbf{DFF} & \textbf{Number} & \textbf{DFF} & \textbf{Number} \\ 
\hline
Gaussian Distribution & 64 & Log-Normal Distribution  & 14 \\ 
Gaussian Mixture Distribution & 33 & Exponential Distribution & 24 \\ 
Not-Fitting & 44 & & \\ 
\hline
\end{tabular}
\end{table}

\vspace{-1cm}

\vspace{-0.1cm}
\subsubsection{Road Network Matching Accuracy}
\begin{figure}[htbp] 
\vspace{-0.5cm}
    \centering
    \begin{subfigure}[b]{0.4\textwidth}  
        \centering
        \includegraphics[width=\textwidth]{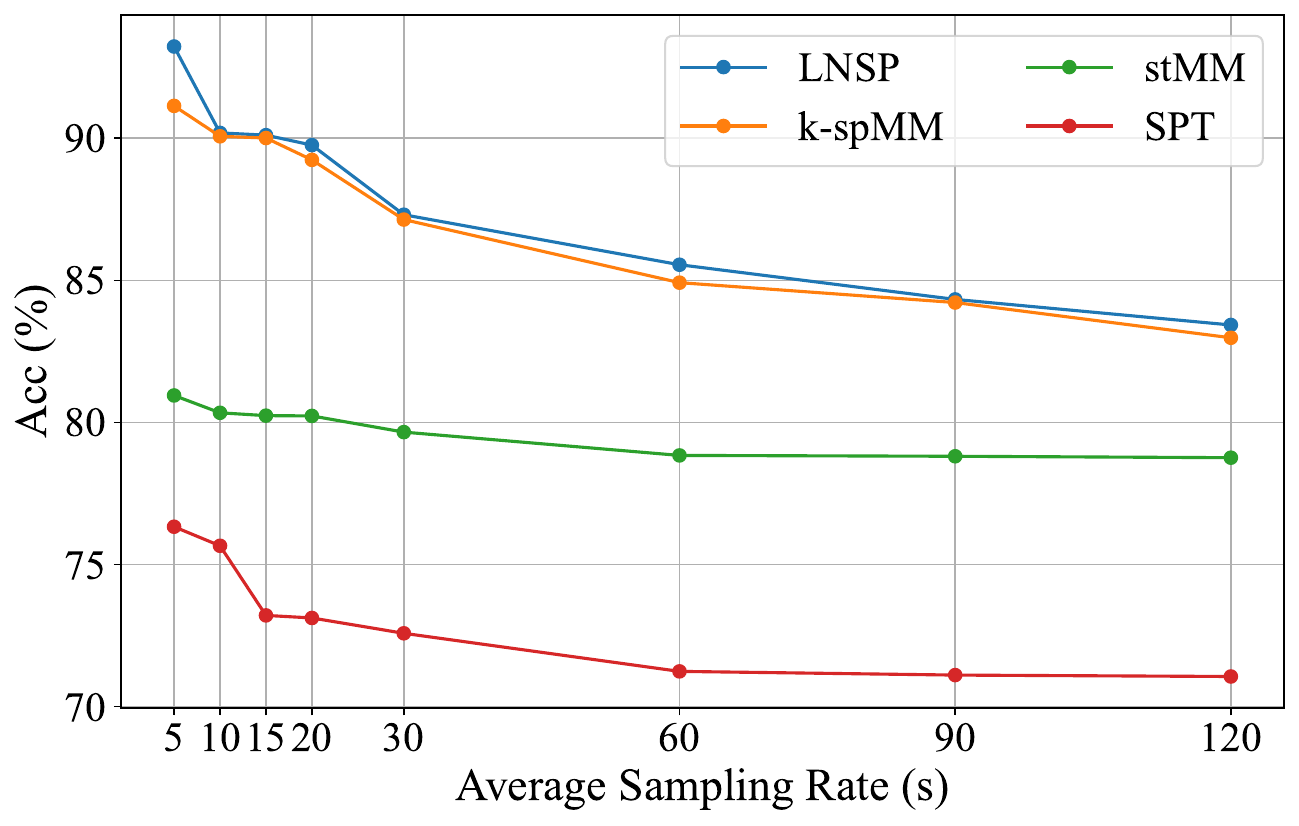}
        \caption{}
        \label{fig:bus_matching_accuracy1Acc}
    \end{subfigure}
    \hfill
    \begin{subfigure}[b]{0.4\textwidth}
        \centering
        \includegraphics[width=\textwidth]{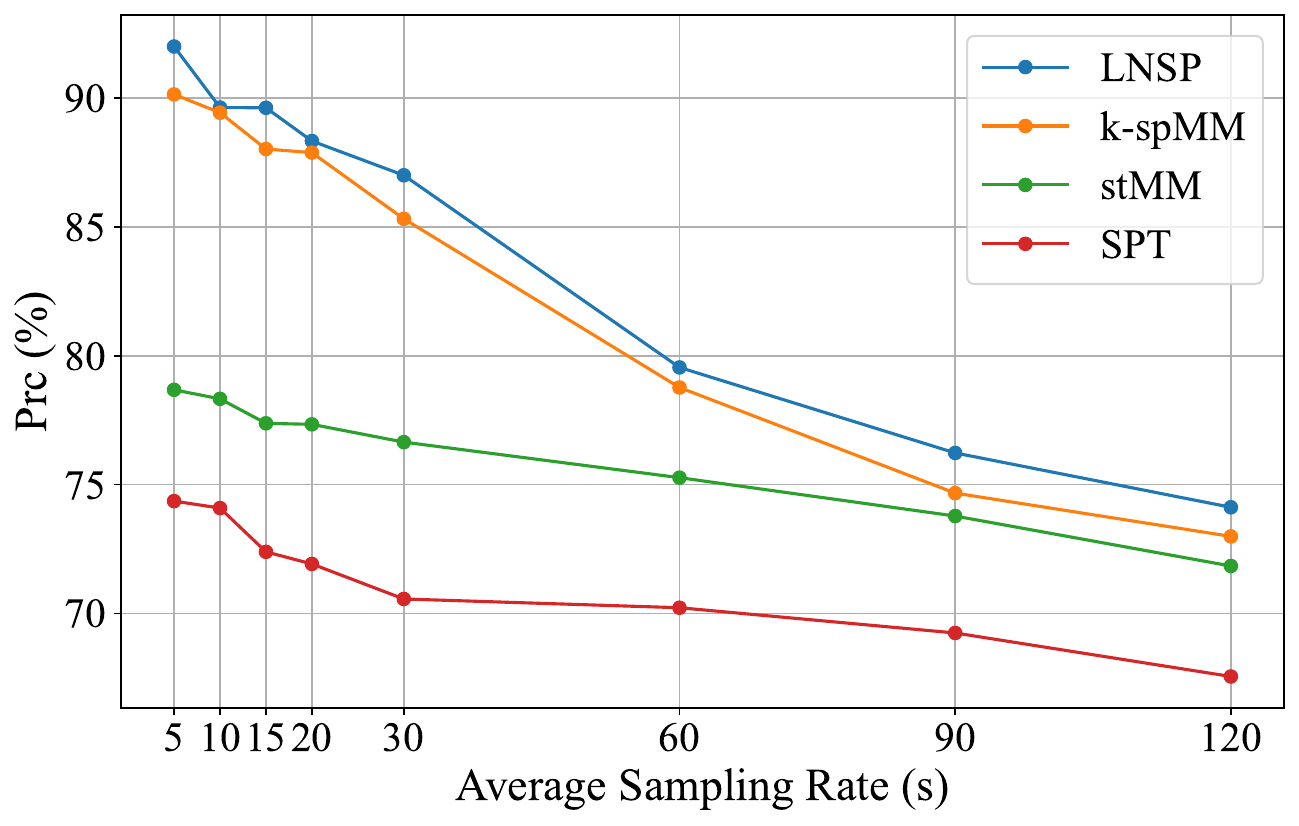}
        \caption{}
        \label{fig:bus_matching_accuracy2Prc}
    \end{subfigure}
    \vspace{-0.3cm}
    \caption{ Overall Performance for  Bus Trajectories}
    \label{fig:bus_matching_accuracy}

\end{figure}

All methods are evaluated on both bus and taxi trajectories under various GPS sampling rates ranging from 10 to 120 seconds, as illustrated in Fig.~\ref{fig:bus_matching_accuracy} and Fig.~\ref{fig:taxi_matching_accuracy}.

The {\SYS} significantly outperforms the SPT~\cite{RAHMANI201341} and stMM~\cite{QUDDUS2015328} algorithms. For example, the accuracy (ACC) for bus trajectories improves by 12.28\%  and 16.9\%  compared to SPT and stMM, with precision metrics increasing by 13.33\%  and 17.65\%  at a 5-second sampling rate.  This advantage is due to the fact that vehicle trajectories do not always follow the global shortest paths. Compared to k-spMM~\cite{10.1145/3191801.3191812}, {\SYS} also performs better at high sampling rates. At a 5-second average sampling rate, accuracy and precision improve by 2.09\% and 1.86\% , respectively. k-spMM struggles in areas with significant GPS errors, where the candidate paths may not be unique, and trajectory curves can deviate greatly from the actual path, leading to incorrect matches.

\begin{figure*}[htbp]
\vspace{-0.6cm}
\centering
\includegraphics[width=0.8\textwidth]{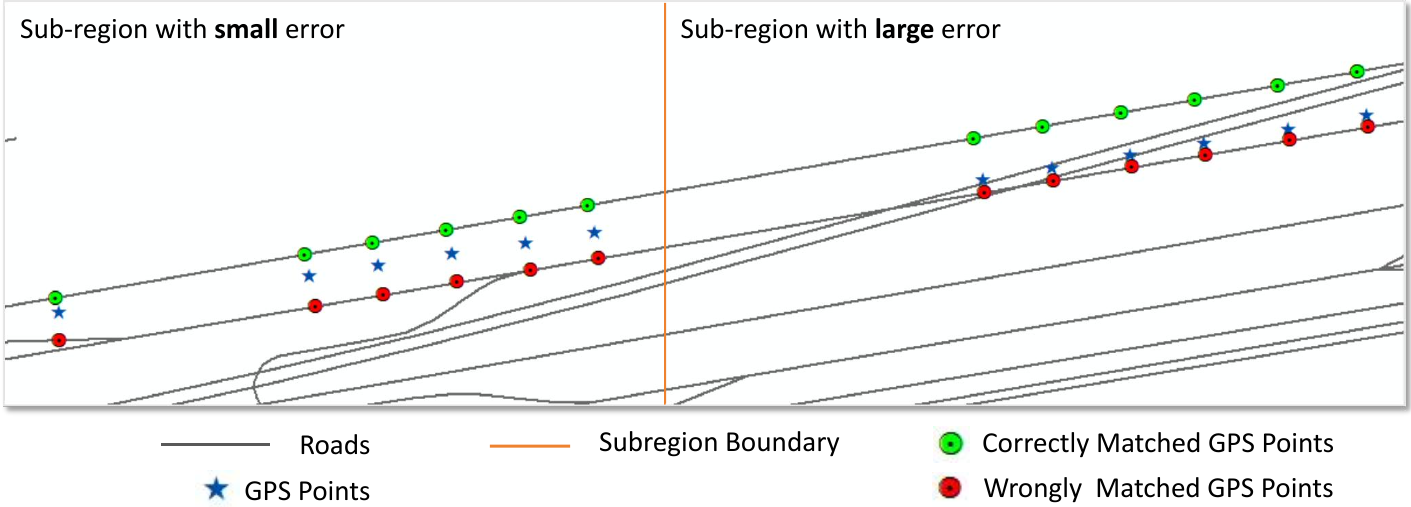}
\caption{Example of Incorrect Matching with the k-spMM Model When Encountering Consecutive Large Errors}
\label{fig:k_spMM_wrong_matching}
\vspace{-0.8cm}
\end{figure*}

As shown in Fig.~\ref{fig:k_spMM_wrong_matching}, the trajectory in the right half is located in an area with significant errors, resulting in a smaller Fréchet distance between the incorrectly matched path and the actual trajectory. Unlike Fréchet distance, our method  {\SYS}  more accurately describes the likelihood of errors between the trajectory and the matching path by calculating the total weight of the LED probabilities for each matching point's subregion along the path. In the case shown in  Fig.~\ref{fig:k_spMM_wrong_matching}, the {\SYS}  model improves the matching by recognizing the error distribution of the region, increasing the weights of the correct candidate points in both the left and right halves, thereby successfully matching the correct trajectory path. Additionally, the algorithm smooths the connections between the matching paths of each local trajectory segment using a sliding window, strengthening the relationship between local and global trajectories. As a result, the {\SYS}  model demonstrates higher matching accuracy.

However, in the taxi trajectory matching results presented in Fig.~\ref{fig:taxi_matching_accuracy}, the matching accuracy of {\SYS} shows slight degradation at medium to low sampling rates. This is due to the more complex road environments for taxis compared to buses, and the GPS error distribution for some taxi data, especially those operating in non-bus routes, is estimated through interpolation. Despite this, {\SYS} still maintains an advantage at high sampling rates.

\vspace{-0.5cm}
\subsubsection{Matching Efficiency}

We conducted experiments to measure the matching time of different methods on the same computer configuration. The results in Table~\ref{tab:comparison of Mt} show that the {\SYS} model has slightly lower matching efficiency than SPT~\cite{RAHMANI201341} and stMM~\cite{QUDDUS2015328} due to its more complex local non-shortest path detection and correction, which require additional computational resources (further explained in the next subsection). However, {\SYS} outperforms the k-spMM model, which uses time-consuming dynamic programming to search for optimal shortest path combinations in complex road networks~\cite{Chao2019ASO}. In contrast, {\SYS} first matches the shortest path and then detects errors and segments potential non-shortest paths based on maximum error points, leveraging the geometric and topological features of the road network. This results in faster and more efficient trajectory segmentation, reducing matching time. Additionally, {\SYS} dynamically adjusts the candidate search range based on error distribution, reducing the number of candidate points compared to k-spMM’s fixed search radius. 
\begin{table}[htbp]
\vspace{-1cm}
\caption{Comparison of Model Matching Time Metric (Mt)}
\label{tab:comparison of Mt}
\centering
\begin{tabular}{c| c| c}
\hline
\textbf{Model} & \textbf{Mt on Bus (ms)}  & \textbf{Mt on Taxi (ms)} \\
\hline
StMM & 116 & 137 \\
SPT & 128 & 125 \\
{\SYS} & 135 & 168 \\
k-spMM & 262 & 272 \\
\hline
\end{tabular}
\vspace{-1.4cm}
\end{table}

\begin{figure}[htbp]
    \centering
    
    \begin{subfigure}[b]{0.45\textwidth}
        \includegraphics[width=\textwidth]{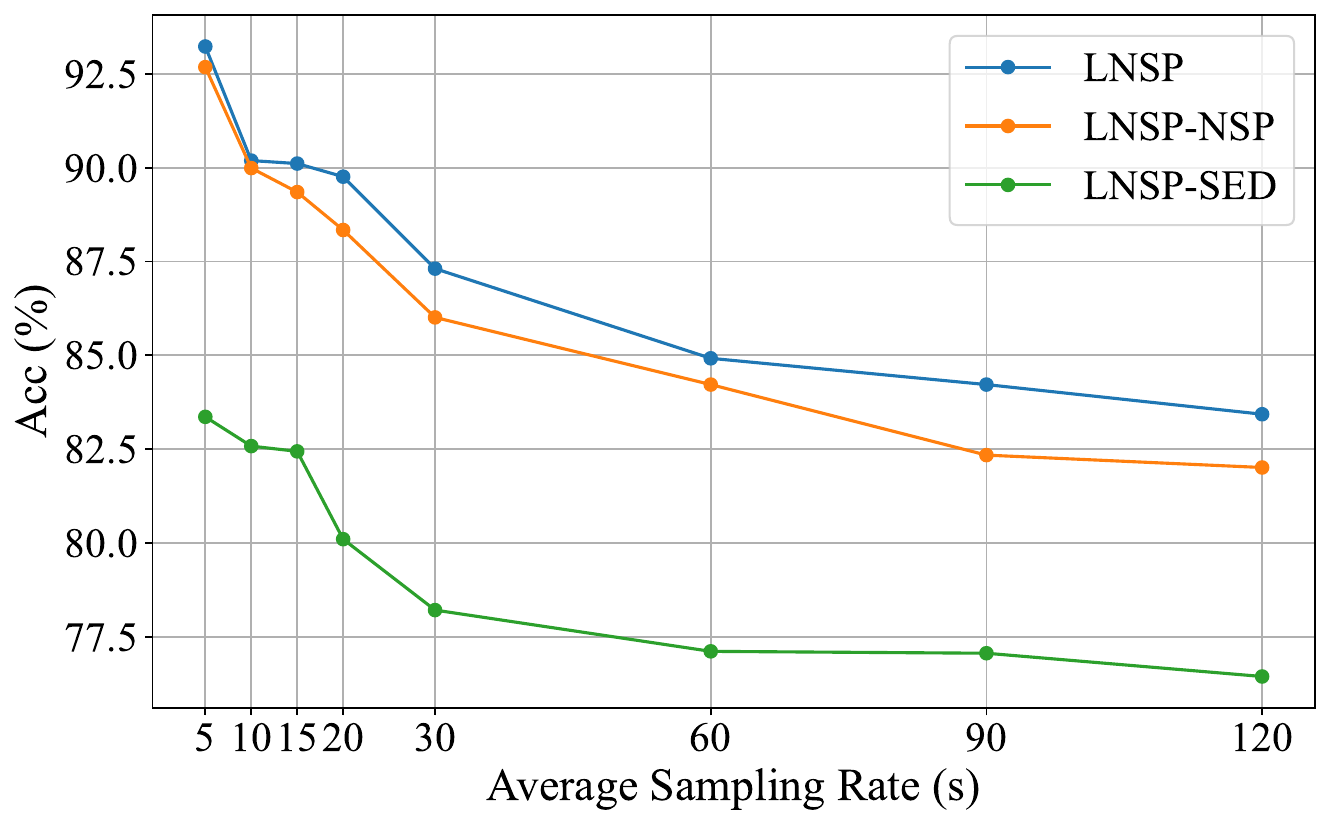}
        \caption{ACC of bus trajectories}
        \label{fig:busaccabl}
    \end{subfigure}
    \hfill 
    \begin{subfigure}[b]{0.45\textwidth}
        \includegraphics[width=\textwidth]{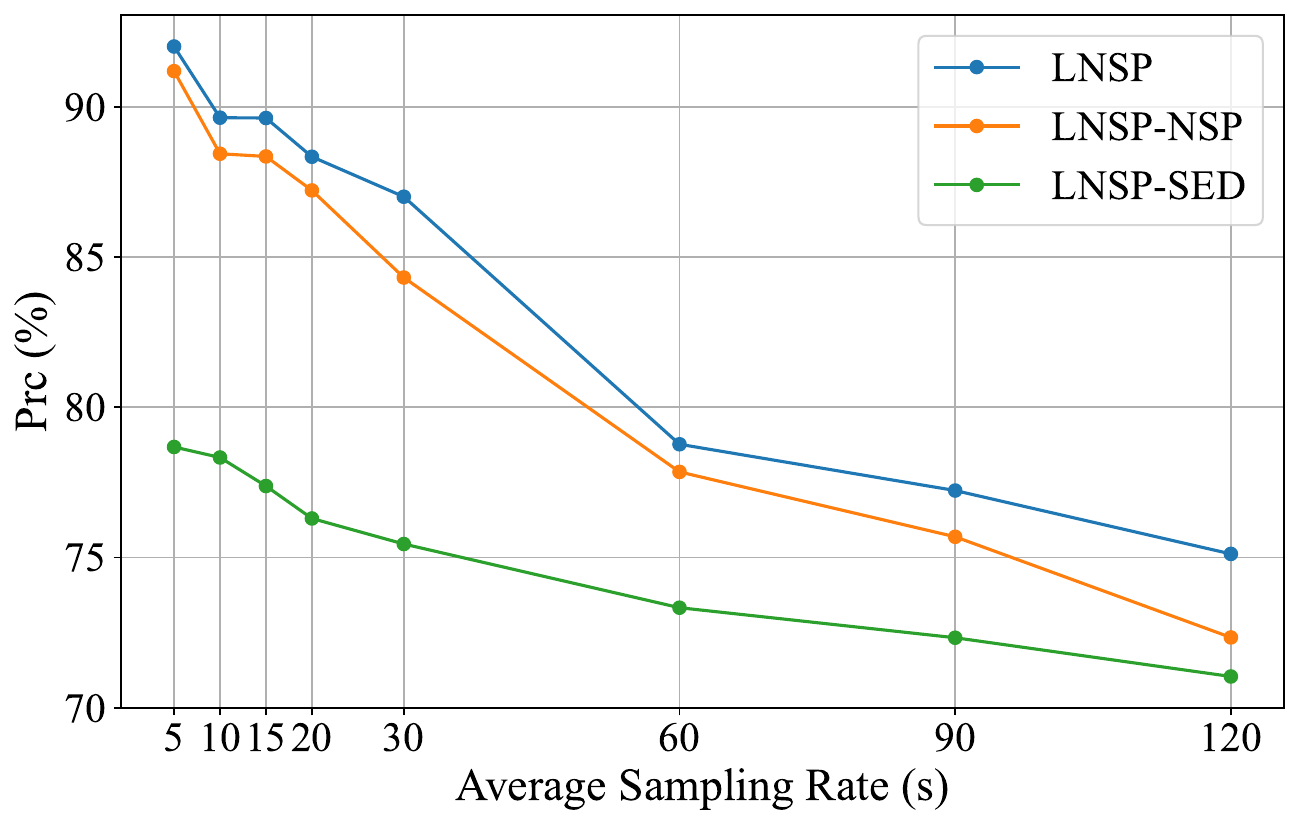}
        \caption{Prc of bus trajectories}
        \label{fig:busprcab}
    \end{subfigure}

    \vspace{0.5cm}

    \begin{subfigure}[b]{0.45\textwidth}
        \includegraphics[width=\textwidth]{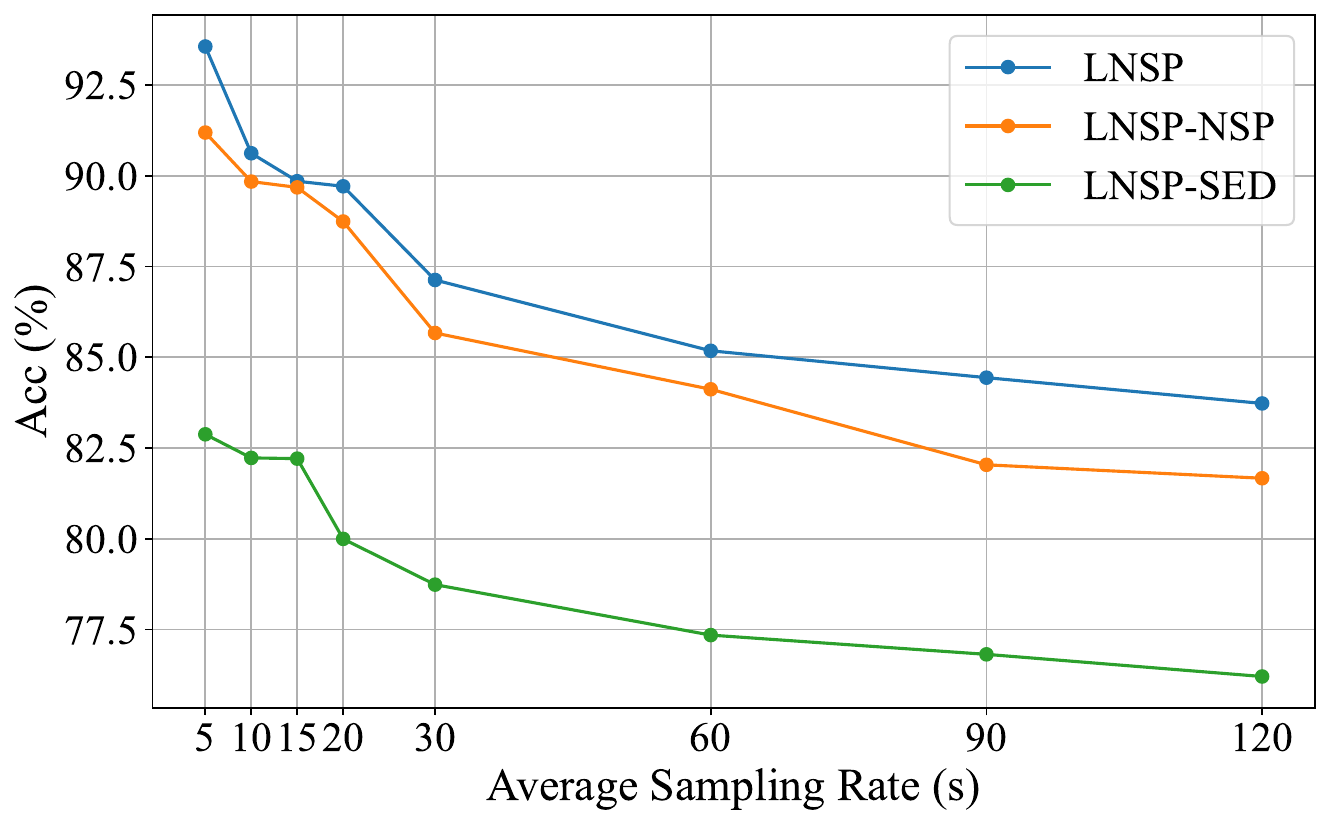}
         \caption{ACC of taxi trajectories}
        \label{fig:taxiaccabl}
    \end{subfigure}
    \hfill 
    \begin{subfigure}[b]{0.45\textwidth}
        \includegraphics[width=\textwidth]{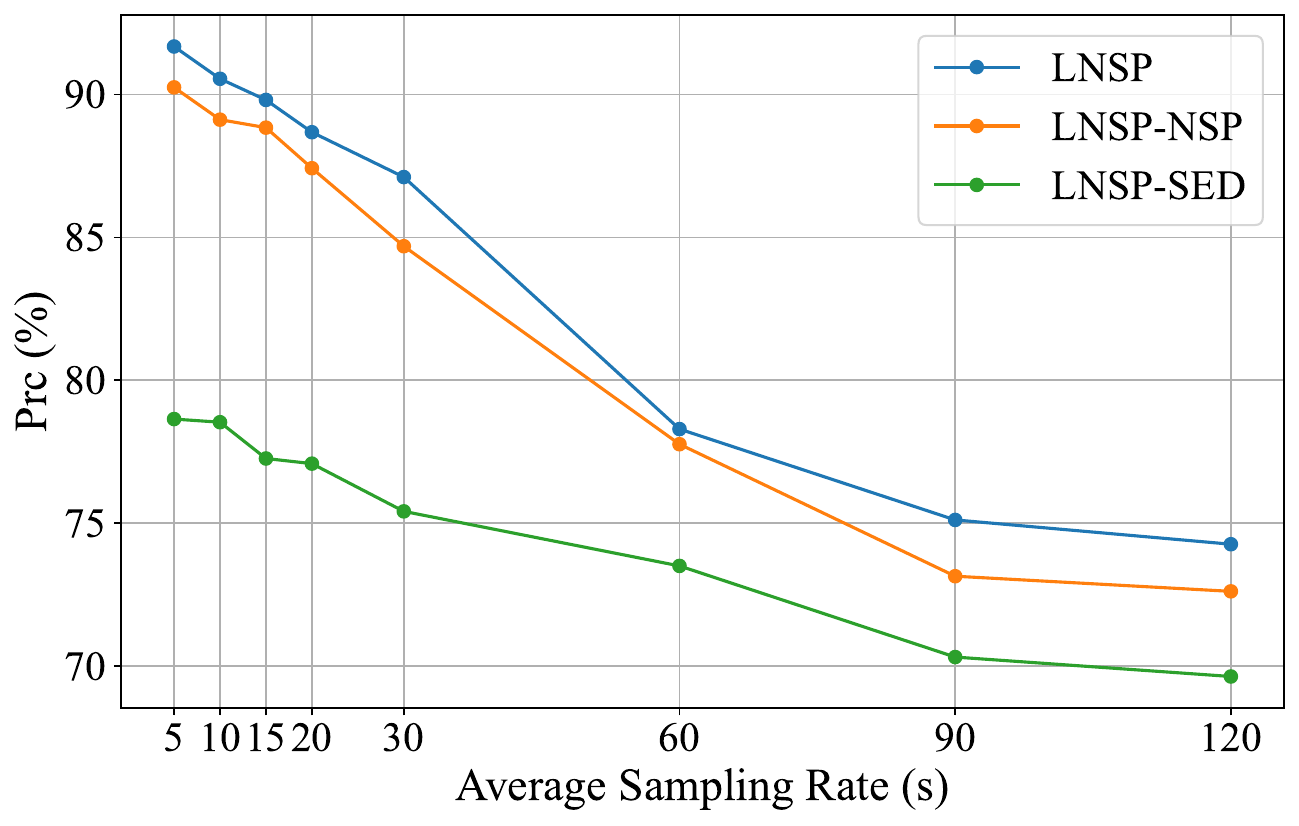}
        \caption{Prc of taxi trajectories}
        \label{fig:taxiPrcabla}
    \end{subfigure}
    \caption{Acc \& Prc of Ablation Models for Taxi/Bus Trajectories}
   
   \vspace{-0.8cm}
\end{figure}

\subsubsection{Ablation Study}
To further analyze the contributions of the sub-region LED-based candidate point search radius setting and path weight scoring (SED) module, as well as the local non-shortest path detection and correction (NSP) module, we conducted experiments by separately removing these modules. The two models {\SYS}$-$NSP (with the SED module removed) and {\SYS}$-$SED (with the NSP module removed) were compared. The matching accuracy metrics are illustrated in Fig.~\ref{fig:busaccabl} and Fig.~\ref{fig:taxiaccabl}. The removal of the local non-shortest path detection leads to a significant decrease in Acc and Prc metrics, indicating that, in the GPS trajectory dataset used in this experiment, deviations from the shortest path are common in local paths. The NSP module's ability to detect and correct these deviations confirms the validity of the approach.

\vspace{-0.4cm}
\section{Conclusion}
\vspace{-0.2cm}
This paper proposes {\SYS}, a novel offline map matching method that effectively addresses key limitations in handling localization errors and path scoring for sparse trajectory data. By leveraging bus trajectories with fixed routes, {\SYS} provides a more accurate, region-specific model for localization error distribution (LED), which enhances the efficiency of the path search process. Additionally, the use of a sliding window technique and region-dependent LED scoring significantly improves matching accuracy, particularly in urban environments with varying geographical factors. Experimental results using real-world bus and taxi datasets from Shenzhen show that {\SYS} outperforms existing methods in both efficiency and accuracy, offering a robust solution for intelligent transportation systems. Future work may explore extending our method under the framework of multimodal unified representation~\cite{xia2023achieving,huang2024unlocking,huang2025semantic}, allowing for more expressive integration of visual, textual, and sensor-based modalities.

\vspace{-0.2cm}
\begin{credits}
\subsubsection{\ackname} This work is supported by Guangdong Basic and Applied Basic Research Foundation (No.2023B1515130002), Shenzhen Industrial Application Projects of undertaking the National key R \& D Program of China (No.CJGJZD20210408091600002), Consultation Research Projects of Guangdong Institute of Engineering Science and Technology Development Strategy(No.2024-GD-5-02) 

\end{credits}
\vspace{-0.5cm}
\bibliographystyle{splncs04_unsorted}
\bibliography{mainref}

\end{sloppypar}
\end{document}